\documentclass[submission,copyright,creativecommons]{eptcs}

\usepackage{iftex}

\ifpdf
  \usepackage[T1]{fontenc}        
\else
  \usepackage{breakurl}           
\fi

\title{WebPie\\ A Tiny Slice of Dependent Type}

\author{Christophe Scholliers
\institute{Department of Mathematics, Applied Statistics and Informatics \\Ghent University}
\email{christophe.scholliers@ugent.be}
}
\def\titlerunning{WebPie: A Tiny Slice of Dependent Types}
\def\authorrunning{Christophe Scholliers}

\usepackage[utf8]{inputenc}

\usepackage{amsmath}

\usepackage{verbatim,fleqn}

\usepackage{url}

\usepackage{graphicx}				
								
\usepackage{listingsutf8}

\protected\def\imp{\ensuremath{^\supset}}
\protected\def\ar{\ensuremath{^\to}}

\DeclareUnicodeCharacter{07CA}{\imp}
\DeclareUnicodeCharacter{07C6}{\ar}
\DeclareUnicodeCharacter{00AC}{ noe }

\usepackage{newunicodechar}
\newunicodechar{¬}{ note t}

\usepackage{color}
\definecolor{lightgray}{rgb}{.9,.9,.9}
\definecolor{darkgray}{rgb}{.4,.4,.4}
\definecolor{purple}{rgb}{0.62,0.31,0.90}

\lstdefinelanguage{JavaScript}{
  keywords={def,Axiom,Inductive,Type,Set,Prop,typeof, new, false, catch, function, return, null, catch, switch, var, if, in, while, do, else, case, break, match, with},
  keywordstyle=\color{blue}\bfseries,
  ndkeywords={class, export, boolean, throw, implements, import, this},
  ndkeywordstyle=\color{darkgray}\bfseries,
  identifierstyle=\color{black},
  sensitive=false,
  comment=[l]{//},
  morecomment=[s]{/*}{*/},
  morecomment=[l]{\%},
  commentstyle=\color{purple}\ttfamily,
  stringstyle=\color{red}\ttfamily,
  morestring=[b]"
}

\usepackage{collectbox}	
\usepackage{amssymb}
\newcommand{\createbox}{%
    \collectbox{%
        \setlength{\fboxsep}{5pt}%
        \fbox{\BOXCONTENT}%
    }%
}
\usepackage {mathpartir}
\usepackage{hyperref}

\begin{document}

\title{WebPie \\ A Tiny Slice of Dependent Typing}

\maketitle

\begin{abstract}
Dependently typed programming languages have become increasingly relevant in recent years.  
They have been adopted in industrial strength programming languages and have been extremely successful as the basis for theorem provers.
There are however, very few entry level introductions to the theory of language constructs for dependently typed languages, and even less sources on didactical implementations. 
In this paper, we present a small dependently typed programming language called WebPie. 
The main features of the language are inductive types, recursion and case matching. 
While none of these features are new, we believe this article can provide a step forward towards the understanding and systematic construction of dependently typed languages for researchers new to dependent types. 
\end{abstract}

\section{Introduction}
Dependent types are an interesting and fascinating subject with many recent developments~\cite{Weirich17,agda08,Coq,deBruijn1983,avron92,harper93,JFP:9060502}. 
Many theorem provers such as Coq~\cite{Coq} and Agda~\cite{agda08} make use of some form of dependent types. 
Unfortunately, the theory and implementation of full fledged systems such as Coq and Agda are complex and consists of thousands of lines of code. 
On the other hand, entry level textbooks on dependent types only give an overview of a core calculus omitting crucial aspects such as inductive types, dependent case matching and recursion. This article aims to bridge the gap between entry level textbooks on types such as ``Types and Programming Languages''~\cite{pierce2002} and the theory of full fledged theorem provers based on dependent typing. While this article provides an entry level explanation on dependent types readers are expected to be familiar with basic type theory and functional programming~\footnote{In particular, readers are expected to be familiar with basic type systems as described in the book "Types and Programming languages"~\cite{pierce2002} up to System-F-$\omega$. Furthermore,  knowledge of basic functional programming language constructs~\cite{Hutton16} such as algebraic datatypes is expected.   }.


The explanation of the dependently typed language constructs and their implementation is done in two stages. In a first stage, we present
WebLF, a very basic dependently typed programming language inspired by the logical framework~\cite{harper93}. 
In a second stage, we add inductive types and motivate how the programmer benefits from them. 
The resulting WebPie language is available online~\footnote{\href{https://users.ugent.be/~chscholl/WebPie/}{https://users.ugent.be/$\sim$chscholl/WebPie/}} with several proof examples showing that the language can be used for small proofs. 

The WebPie language is conceived as a language laboratory to further investigate the language features that are
present in dependent type theory. The current implementation does not feature support for advanced features such as tactics for interactive theorem proving. 
By making these simplifications we hope to hit a sweet spot where the reader can be inspired to understand dependent typing and how to implement their own dependently typed programming
language without being overwhelmed with the difficulties.




\section{Getting Familiar with Dependencies}
Often dependent types are perceived as exotic and difficult. There is however, no reason why dependent types are any more difficult than other concepts in computer science. 
The idea of dependency is in fact pervasive in many elementary programming language constructs. 
When using these constructs most programmers do not consciously think about the dependencies they are making. 
As a stepping stone towards dependent types it is worthwhile to make these dependencies in basic language constructs explicit. 
Let us take a look at three examples which should be familiar to most programmers.

\textbf{Pure functions} define dependencies between their arguments and return values. 
For example, the \texttt{increment} function when called with the argument 3 will return the value 4, which of course depends on the given argument 3. 
Looking at the types and values involved, functions capture a \emph{dependency from terms to terms}. 


\textbf{Generics}
A second (distinct) form of dependency is encountered
when making use of generics in programming
languages such as Java~\cite{Java}. When making a new container class
(for example  \lstinline[language=Java]{new Vector<Integer>())}, the programmer explicitly
passes a type to the container to receive a constructor
that creates instances of a new type of container. 

Important to note, is the dependency between the constructor and the given type.
Similarly to how pure functions represent terms which dependent on other terms, generics encode \emph{terms which depend on types}. 
This also means that in order for a system like this to exist that suddenly the distinction between types and terms are starting to blur. 
In systems without generics there is no way for the programmer to compute with types. 
They only serve as annotations. In this system however, the programmer can use types as if they would be terms.
The use of types in this example is however, limited to passing types as arguments to create terms.

\textbf{Type constructors}
In a system with generics as described before, we can only make a  \lstinline[language=Java]{Vector<Integer>} constructor,
but we still do not have the power to protect functions to
differentiate between a vector of \lstinline{Integers} or of \lstinline{Strings}.
Also, at the type level, we need to be able to create new types.
It would be silly/impossible to create a table with all possible
types we would ever need in our programs. Therefore, many
languages introduce the concept of a type level abstraction.
In such a system, we can create functions which create other
types. For example,\lstinline[language=Java]{Vector<Integer>} is a function which
receives a type and returns the type “Vector which contains
Integers”. This means that type constructors capture a dependency
from types to types.

\begin{table}[h!]
\centering
\begin{tabular}{ll}
Name &  Dependency  \\
\hline
Function &  Term $\mapsto$  Term  \\
Generics &  Type $\mapsto$ Term \\
Type Abstraction &   Type $\mapsto$  Type \\  
\end{tabular}
\caption{Type dependency overview.}
\label{overview}
\end{table}

Table~\ref{overview} shows the different kinds of dependencies observed
in the three examples. Attentive readers may have noticed
there is one dependency missing from the table (terms to
types). 
The idea of dependent types is to create a system
that captures a dependency from \emph{terms to types}. In the next
section, we give an example of why such a system could be
useful. In the remainder of the paper, we give an overview of
how to implementation a tiny dependently typed language
in JavaScript.

\section{Dependent Types by Example}
As it might not be obvious why dependent types might be useful we give a small practical example. 
After this practical example we look into how dependent types can be used as a proof system. 

In many languages there is a function which allows the programmer to print a number of arguments by means of a format string. 
This format string indicates what the type of the arguments of the rest of the arguments should be. 
For example in C the following piece of code shows how the \lstinline{printf} function is applied to two arguments. 
The first argument is the format string in which, \lstinline[language=C]{
As expected executing this piece of code will print {\tt "The number is 10"}.  
\begin{lstlisting}
printf("The number is %d", 10);
\end{lstlisting}

Unfortunately, the type of the arguments after the format string are not type checked and thus can lead to erroneous behaviour at runtime.
In the context of dependent types the crucial insight is that the type of the \lstinline{printf} function \textit{depends} on the  format string (value) given as an argument.
For example, the following two invocations to the printf functions have a different type. 
\begin{lstlisting}
printf :: String -> Int -> Void
printf("The number is %d",  ... );

printf :: String -> String -> Void
printf("The string is %s", ... )
\end{lstlisting}

In a dependently typed programming language it is possible to define \lstinline{printf} as a dependently typed function whose range type depends on the format string it receives. 
The definition of such a function is shown below.

\begin{lstlisting}
def printf(s : String) : (computeType s) {
	% create a function based on the format string
}
\end{lstlisting}

The syntax \lstinline{(s: String)} makes the argument \lstinline{s} accessible in the rest of the type signature. 
In our case the range of the type is computed by a function \lstinline{computeType}. 
The type checker is able to verify that for any possible format string, the \lstinline{computeType} function correctly computes the type of the function returned by the  \lstinline{printf} function. 
The type of the \lstinline{printf} function is a \emph{dependent type} which is typically defined using the \lstinline{Π} symbol. 
The type of the \lstinline{printf} function is thus \lstinline[language=C]{Πs:String.(computeType s)}. 
Note that, $\Pi$, generalises traditional function types when the arguments of the dependent type are not used in the rest of the function signature. 
For example, the type \lstinline[language=C]{Πa:Int.String} is equivalent to \lstinline[language=C]{Int->String}.


\section{WebLF: Dependent Types for Proof Assistance}
\label{sec:weblf:taste}
WebLF is an implementation of a logical framework~\cite{harper93}. 
In a logical framework the developer can define the syntax, rules and proofs of formal systems by means of a dependently typed lambda calculus. In order to prove propositions, a logical framework exploits the correspondence between computer programs and mathematical proofs.  This correspondence known as the Curry-Howard correspondence\cite{Howard1980}, states that there is an isomorphism between \emph{types and propositions}  and \emph{proofs and programs}. For example, logical implication \lstinline{⊃} corresponds to function type $\rightarrow$. Universal quantification $\forall$ corresponds to the dependent function type \lstinline{Π}. 
Given the Curry-Howard correspondence stating propositions corresponds to defining a type while giving a proof consists of giving an implementation for the given type. 

In WebLF, defining the syntax of the formal system consist of stating axioms to define what the objects of the formal system are and how to construct them. Defining the semantics of the formal system consist of defining a set of judgements over these objects. 
Once the syntax and semantics of the formal system are defined, the programmer can use WebLF to prove propositions over the system.  

In WebLF, there are two essential constructs \lstinline{Axiom} and \lstinline{def}. 
As the name indicates, \lstinline{Axiom} allows the programmer to postulate assumptions which are taken for granted without giving an explicit proof for them.  For example, to state that there is a set of natural numbers the programmer can state:  \lstinline{Axiom Nat : Set}. Here \lstinline{Set} denotes the type of proper types, i.e. \lstinline{Set} corresponds to kind star in Haskell.

With the \lstinline{def} construct the programmer can state and prove propositions following the Curry-Howard correspondence. 
The syntax of \lstinline{def} follows the syntax of ordinary function definitions found in many programming languages. 
To make this more concrete let us look at two small examples of how to encode formal systems in WebLF. 
The first example consist of encoding first-order logic, the second example encodes the natural numbers (with induction)~\footnote{More elaborate examples are available online.}.

\subsection{First-Order Logic}

In order to model first-order logic in WebLF we first declare a new set to represent the objects (formulas) of the logic (as shown in figure \ref{fig:weblf:syntax:semantics}).
This is done by adding an axiom stating that there is a set \lstinline{o}. 
The type of the set of objects itself is \lstinline{Set}. 

\begin{figure}[thp] 
\begin{lstlisting}[inputencoding=utf8]
% Syntax 
% o are propositions
Axiom o : Set;
Axiom ⊃ : o → o → o;

% Judgment 
Axiom true : o → Set;
%       ϕ
%       .
%       ψ
%  ----------- [ Imp-I ]
%    ϕ ⊃ ψ
Axiom impI : Πp:o.Πq:o. 
             ((true p) → (true q)) → 
             (true (⊃ p q));
\end{lstlisting}    
\caption{WebLF: Syntax and semantics of FOL.}
\label{fig:weblf:syntax:semantics}
\end{figure}  

Then axioms for each of the constructors of \lstinline{o} are added. 
Here we only show the constructor for implication $\supset$. 
Its type indicates that $\supset$ expects two objects and that it returns a new object.
If \lstinline{p} and \lstinline{q} are two objects, \lstinline{ (⊃  p q) } is also an object. 

Subsequently, we define what it means for an object of the logic to be true. 
This is done by stating that there is a judgment \lstinline{true} over objects.
Encoding a judgement in WebLF is done by defining a set which is parametrised over the defined objects. 

Then a single introduction rule, implication introduction, is defined. 
In order to express the type of this introduction rule we need a dependent type. 
In WebLF dependent types are written as \lstinline{Πn:T.X} where \lstinline{X} might refer to the variable \lstinline{n}. 
If \lstinline{X} does not refer to \lstinline{n} the type can be written more traditionally as \lstinline{T→X}.

The type of the introduction rule \lstinline{impI} thus denotes that the introduction rule is a function expecting two objects ( \lstinline{p} and \lstinline{q}) and a function which can transform any program \lstinline{(true p)}
 into a program \lstinline{(true q)}. Given these arguments \lstinline{impI} returns a program \lstinline{(true (⊃  p q))}.

Given these definitions we have enough axioms to define and prove our first proposition. 
In Figure~\ref{fig:weblf:proof:fol} a function \lstinline{imp_a_b_a} is defined. 
Through the Curry-Howard correspondence between \lstinline{Π} and $\forall$ its type signature \lstinline{ΠA:o.ΠB:o.(true (⊃  A (⊃  B A)))} encodes the proposition $\forall A,B.A \supset B \supset A$.
The term in the body of the function is the program/proof for this type/proposition. 
It is built by a nested application of the \lstinline{impI} rule. 
The outer application of  \lstinline{impI} binds the variable \lstinline{p} to \lstinline{A} and the variable \lstinline{q} to \lstinline{(⊃ B A)}. 
The third argument to \lstinline{impI} thus must be a function with the following signature \lstinline{(true A) → (true (⊃  B  A))}.
This function is constructed on lines 4-5. The argument \lstinline{a} of the function can be used to construct a proof for \lstinline{ (true (⊃  B  A))} by a second application of the   \lstinline{impI} rule.  In this second application the variable \lstinline{p} is bound to \lstinline{B} and the variable \lstinline{q} to \lstinline{A}. We therefore, need to supply a function with the following signature  \lstinline{(true B) → (true A)}. This is quite easy because the variable \lstinline{a} is exactly \lstinline{(true A)}. It is thus sufficient to return the variable a. The inner application of \lstinline{impI} thus returns us a proof of \lstinline{(true (⊃  B A))}. Therefore,  the outer application of \lstinline{impI} returns us a proof  \lstinline{(true (⊃  A (⊃  B A)))}, exactly what we needed. 

\begin{figure}[thp] 
\begin{lstlisting}[inputencoding=utf8,xleftmargin=2em]
%  |- A ⊃ B ⊃ A.
def imp_a_b_a(A:o,B:o) : (true(⊃ A (⊃ B A))) {
  (impI A (⊃ B A)
        (λa:(true A).
            (impI B A (λb:(true B).a)))) 
}
\end{lstlisting}    
\caption{WebLF: Proof example in the domain of FOL.}
\label{fig:weblf:proof:fol}
\end{figure}  

%
%

At his point it is worth noting that there are no guarantees provided by WebLF (and other proof assistants) that the encoded formal system corresponds to the actual system. 

\subsection{Natural Numbers and Induction}
In most cases simply applying the set of constructor axioms will not be sufficient to prove more interesting propositions. 
For example, to prove interesting properties over the natural numbers we will often need induction as a reasoning principle. 
In WebLF such reasoning principles can be added as an axiom. 
Such an encoding is shown in Figure~\ref{fig:weblf:induction}.

\begin{figure}[h!] 
\label{fig:weblfnat}
\begin{lstlisting}[inputencoding=utf8]
Axiom Nat : Set;
Axiom z   : Nat;
Axiom s   : Nat->Nat;

% Adding two numbers
Axiom plus      : Nat -> Nat -> Nat ->Set;
Axiom plus_zero : Πx:Nat.(plus x z x);
Axiom plus_x_y  : Πx:Nat.Πy:Nat.Πz:Nat.
                 (plus x y z) -> 
                 (plus x (s y) (s z)); 

% Induction principle for Nat
Axiom nat_ind   : ΠP:Nat->Set.
                  Πp_zero:(P z). 
                  Πp_succ:(Πx:Nat.Πp:(P x).
                                     (P (s x))). 
                  Πx:Nat.(P x);

def plus_zero_x( x:Nat ) : (plus z x x) {
	((nat_ind (λx:Nat.(plus z x x))
	  (plus_zero z)
	  (λx:Nat.  
	   (λIH:(plus z x x).  
	    (plus_x_y z x x IH)))) 
	 x) 
}
\end{lstlisting}   
\caption{WebLF: Induction and natural numbers.}
\label{fig:weblf:induction}
\end{figure}

First the Set \lstinline{Nat} is defined with two constructors \lstinline{z,s} denoting zero and successor.
Then a judgement \lstinline{plus} is defined which relates three numbers where the third number is the sum of the first two. 
Two rules for this judgement are defined. The first, \lstinline{plus_zero}, encodes that adding \lstinline{z} to any number \lstinline{x} is simply \lstinline{x}. The second, \lstinline{plus_x_y}, encodes that when given a proof \lstinline{(plus x y z)} it can be deduced that \lstinline{(plus x (s y) (s z))}.  

Subsequently the induction principle over \lstinline{Nat}, \lstinline{nat_ind}, is defined.  
This encoding states that for all propositions \lstinline{P} over \lstinline{Nat}, when given a proof for that proposition for \lstinline{z} and a function \lstinline{p_succ} that when given any number \lstinline{x} and a proof of \lstinline{P x} gives a proof of \lstinline{(P (s x))} we can derive that the proposition holds for all elements in \lstinline{Nat}.  

Having defined \lstinline{Nat}, \lstinline{plus} and \lstinline{nat_ind} we can state and prove that for all \lstinline{Nat} we can find a proof that \lstinline{(plus z x x)}.
In spite of being a bit verbose the proof \lstinline{plus_zero_x} follows standard reasoning by induction. 

\section{A Formal Introduction to WebLF}
Now that we have given an overview of how to work with WebLF, we will give an overview of the formalisation of WebLF.  
\subsection{Syntax}
Figure~\ref{fig:entities} shows the syntax of WebLF.
An expression is either a variable, a type universe, a lambda abstraction, a $\Pi$ type or an application. 
Type universes are needed to denote the types of types, and the types of types of types and so on. 
The index $i$ in a type universe is a number indicating the universe level. 
\begin{figure}[h!]
\center
  \begin{tabular}{ r l l r }
  	$e $ & $::=$ & & $$\\ 
	 & $|$ & $x$  & $variable$\\
	 & $|$ & ${\tt Type}_{i}$  & $sorts$\\
	 & $|$ & $\lambda x:e.e$  & \textit{abstraction}\\
     & $|$ & $\Pi x: e.e$  & \textit{abstraction}\\
     & $|$ & $e$ $e$  & $application$ \\
	 & $|$ & ${\tt Axiom}~x:e$  & \textit{Axiom}\\
\end{tabular}
  \caption{WebLF Expressions.}
  \label{fig:entities}
\end{figure} 
 
\subsection{Type Rules}
The typing rules for our dependently typed lambda calculus are shown in Figure~\ref{fig:Typing}. The type of a variable is looked up in the context $\Gamma$.  The type of an abstraction is a $\Pi$ type but only when the expression denoting the type of the function $e_t$ has type ${\tt Type_i}$.  The type of a $\Pi$ type is a universe where the level is the maximum of the universe types of the argument and the body. The type of a universe is the universe type where the level is incremented by one. Finally, the type of an application is obtained by substituting the term $t_2$ in the return type $e_r$ of the function type. This is only correct when the argument type is equal to the expected type $e_t = e_a$. This equivalence is defined by normalisation as shown in the next section.  
 
 \begin{figure}[h!] 
\createbox{ $\Gamma \vdash  x : $T}
\begin{mathpar}
 
 \inferrule*[Right=(T-Var)]
  { x:{\tt T} \in  \Gamma }
  { \Gamma \vdash x : {\tt T} } 
 \hva  \and \and
 \inferrule*[Right=(T-Abs)]
  { \Gamma,x:e_t \vdash e_b : e_{tb}   \\ \Gamma   \vdash e_t : {\tt Type_i}  }
  { \Gamma \vdash  \lambda x:e_t.e_b  : \Pi x:e_t.e_{tb} }

\\
 \inferrule*[Right=(T-PI)]
  {       \Gamma \vdash e_t  : {\tt Type_i}    \\  \Gamma,x:e_t \vdash e_b : {\tt Type_j}    }
  { \Gamma \vdash  \Pi x:e_t.e_{b} :  {\tt Type_{max(i,j)}} }
    \\
 \inferrule*[Left=T-Univ]
  {  }
  { \Gamma \vdash {\tt Type_i} : {\tt Type_{i+1} } }\\

   \inferrule*[Right=(T-App)]
  { \Gamma \vdash t_1: \Pi x:e_t.e_r \\  \Gamma \vdash t_2:  e_a \\ e_t = e_a }
  { \Gamma \vdash t_1\: t_2 :  [x\mapsto t_2] e_r}

\end{mathpar}
  \caption{WebLF typing rules}
  \label{fig:Typing}
\end{figure}

\subsection{Substitution}
Substitution is different from usual substitution in the simply typed lambda calculus as shown in Figure~\ref{fig:substitution}. 
When substituting variables over an abstraction, substitution is performed both at the type level and in the function body. 
Just as in the simply typed lambda calculus, it is important to avoid variable capturing in the rule for abstraction and $\Pi$ types. 
In an implementation it is necessary to perform $\alpha$ renaming in case variables would be captured, or make use of a nameless encoding such as De Bruijn index~\cite{bruijn72}. 
\begin{figure}[h!]
\center
  \begin{tabular}{ l l l r }
$[x\mapsto s]~x $ & $=$ & $s$ & $$\\ 
$[x\mapsto s]~y $ & $=$ & $y$ & $ $\\ 
                               &     & $ y \not= x  $          & \\ 

$[x\mapsto s] ~{\tt Type}_i  $ & $=$ & {\tt Type}$_i$ & $$\\ 
$[x\mapsto s] ~\lambda y:e_t.e_b $ & $=$ & $\lambda y:[x\mapsto s] e_t.[x\mapsto s] e_b $& \\ 
                                   &     & $ y \not= x \wedge y \not\in Fv(s) $          & \\ 
$[x\mapsto s] ~ \Pi y:e_t.e_b$ & $=$ & $\Pi y:[x\mapsto s]e_t.[x\mapsto s]e_b$ &\\ 
                               &     &  $y \not= x \wedge y \not\in Fv(s) $    &\\ 
$[x\mapsto s ] ~ (e_t~e_b) $ & $=$ & $([x\mapsto s ]e_t~[x\mapsto s ]e_b)$ & $$\\ 

\end{tabular}
  \caption{Web$\Pi$ substitution.}
  \label{fig:substitution}
\end{figure}                                            

\subsection{Normalisation and Equivalence}
In the \textsc{T-App} rule the type checker needs to be able to compare two expressions. Equivalence of expressions is defined here by evaluating both expressions and then check whether the evaluated values are syntactically equivalent. Traditional evaluation would leave certain expressions in a form where equivalent terms are not syntactically equivalent. In order to obtain a better approximation we define an alternative evaluation strategy which \emph{normalises} expressions so that equivalent terms are also syntactically equivalent.  
Without normalisation two expressions could be equivalent but have a different syntactic form. 
Figure~\ref{fig:webpie:Normalisation}, in the appendix gives the big step normalisation rules for Web$\Pi$. 
Cases for expression not shown in this figure are already normalised. \\


\section{Implementing WebLF}
In this section we give an overview of the implementation of the WebLF language.
There are two major components to the implementation of WebLF, first the parser transforms the input text into a data structure. 
This data structure is then used by the type checker in order to verify that the provided program is type correct. 
While we try to be as complete as possible, we have deliberately omitted a number of details in our explanation.

First, as we try to focus on the implementation of the dependently typed programming language we do not explain any details about the parser.  

Second, during type checking things might go wrong because the user has supplied the type checker with an erroneous program.  WebLF has a number of places in the type checker to inform the programmer about what might be wrong with the program. 
In the exposition below we omitted the error handling code as we believe it would distract the reader from more important matters. 

\subsection{Representation of Expressions}
In our implementation we make use of adt-simple, a JavaScript library for algebraic datatypes~\footnote{More information on how to install and use the library can be found on: https://www.npmjs.org/package/adt-simple}. 
Datatypes are expressed with the keyword {\tt union}, similar to {\tt data} in Haskell. 
The {\tt deriving} keyword allows certain operations over the algebraic datatypes to be generated by the adt-simple library. 
For example, deriving a function to determine the equality of expressions is done by deriving {\tt adt.Eq}. 
Similarly, we derive the ability to pattern match over an algebraic data type with {\tt adt.Extractor}. 
Finally, deriving a function to print an algebraic data type is done by deriving {\tt adt.ToString}.  
Algebraic datatypes are defined using, JSON syntax. 

In the figure below we define the \lstinline{Expression} datatype.
To keep the implementation simple, variables names are represented with strings. 
Note that in our definition we derive {\tt Eq}, {\tt Extractor} and {\tt ToString}. 
 
\begin{lstlisting}[keywords={union,deriving,function,match,case,return}]{JavaScript}
union Expression {
 Variable {value    : String},
 Universe {value    : Number},
 Lambda   {variable : String,
           type     : Expression,
           body     : Expression},
 Pi       {variable : String,
           type     : Expression,
           body     : Expression},
 App      {fun      : Expression,
           arg      : Expression}
} deriving (adt.Eq,adt.Extractor,adt.ToString)
 \end{lstlisting}

\subsection{The Environment}
In order to keep track of which variables have which types a type checker maintains an environment with variable-type bindings.
The operations on this environment consist of extending the environment with a new binding and looking up the binding of a variable. 
In WebLF such an environment would theoretically be enough, but as WebLF also allows programmers to introduce definitions it not only keeps track of the variable-type bindings but also of variable-value bindings. Apart from this somewhat strange binding relation the environment supports the basic operations one would expect. 
Extending the environment with a variable-type binding: \lstinline{extend_type(var, type, env)}, extend the environment with a type-value binding: \lstinline{extend_type_value(var, value, env)} and operations to lookup values and types \lstinline{lookup_val(var,env)}, \lstinline{lookup_type(var,env)}.

\subsection{Typing Expressions}
The type checker is implemented as a function, \lstinline{type_check}. This function receives two arguments, the context and the expression to type and returns the type of the expression. 
In the code snippet below we see that the function does a case match over the given expression. In case of a variable, the type of the variable is looked up in the context. When it is a universe the type is the universe level plus one. This corresponds to the \textsc{T-Var} and \textsc{T-Univ} rules. The other cases are bit longer and are split up in separate functions. 

\begin{lstlisting}[keywords={function,match,case,return}]{JavaScript}
// Context -> Expression -> Type 
function type_check(ctxt, expr) {
     match expr {
       // ctxt |- x : ?
       case: Variable { value : x } : 
         return  lookup_type(x,ctxt);
       // ctxt |- (Type i) : ?  
       case: Universe { value : i } : 
         return  Universe(i+1);
       case: Lambda{variable:x,type:t,body:b} :
         return check_lam_type(x,t,b,ctxt);  
       //...  
     }
}
\end{lstlisting}

In order to typecheck a lambda expression we first determine the type of the body under the assumption that the variable binding has the given type. In the implementation this is done by extending the context with a variable type binding and recursively calling \lstinline{type_check} (Line 2-3). Then we determine the type of the type declaration itself. In order to be properly typed the type declaration on lambda expression should be a universe type. The resulting type is a $\Pi$ type. 

\begin{figure}[h]
\begin{lstlisting}
// ctxt |- λx:t.b : ?
function check_lam_type(x,t,b,ctxt) {
   var t_body = type_check(extend_type(x,t,ctxt),b);
   var univ_t = type_check(ctxt,t);
   assertUniverse(univ_t,ctxt);
   return Pi(x,t,t_body, b); 
}
\end{lstlisting}
\caption{Type checking lambda abstraction}
\end{figure}

The most interesting type checking rule is the rule for application. We first determine the type of the function \lstinline{f} and the argument \lstinline{a}. The type of the function should be a $\Pi$ type and its argument type \lstinline{typef.type} should be equal to the type of the argument. The resulting type is built by substituting the binding variable of the $\Pi$ type by the expression \lstinline{a} in the body of the $\Pi$ type. 
In order to compare the two expressions we make use of a function \lstinline{check_equal} which first normalises the two expressions and then structurally verifies whether the two expressions are equal. 

\begin{figure}[h]
\begin{lstlisting}
// ctxt |- (f a) : ?
function check_app_type(f,a,ctxt) {
   var typef  = infer_type(ctxt, f);
   var typea  = infer_type(ctxt, a);
   assertPiType(typef,ctxt);
   check_equal(typef.type,typea,ctxt);
   return subst(typef.variable,a,typef.body);
}   
\end{lstlisting}

\caption{Type checking applications}
\end{figure}
   
The type of a $\Pi$ type is determined by first type checking the declared type of the argument. Then the body is type checked under the assumption that the binding has the correct type. Both of these types \lstinline{typet} and \lstinline{typeb} should be universe types. The resulting type is a universe type at level max of the two universe types. 

\begin{figure}[h!]
\begin{lstlisting}{JavaScript}
// ctxt |- Πx:t.b  : ?
function check_pi_type(x,t,b,ctxt) {
   var typet = type_check(ctxt, t); 
   var ectxt = extend_type(x,t,ctxt);
   var typeb = type_check(ectxt, b); 
   assertUniverse(typet,ctxt);
   assertUniverse(typeb,ctxt);
   return Universe(max(typet.value,typeb.value));
} 
\end{lstlisting}
\caption{Type checking $\Pi$ types}
\end{figure}

\subsection{Normalising Expressions}
Normalisation closely follows the formal definition. 
For example, to normalise a lambda abstraction we recursively normalise the type and the body of the lambda.  

\begin{lstlisting}
function normalise(e,ctxt) {
	match e  {
	        //...
			case Lambda { variable:x, type:t, body:b } : 
			   return Lambda(x,normalise(t,ctxt), normalise(b,extend_type(x,t,ctxt)));
			//...
	}
}
\end{lstlisting}

%
%
%

\subsection{Limitations}
Although extremely concise, WebLF forms the basis of many early proof assistant.
It can be of great help by warning the programmer in case there is a reasoning error in his/her proof. 
Once a proof is accepted by the theorem prover the programmer can be confident that all the reasoning steps are sound. One of the limitations of WebLF is that the programmer needs to provide a very large set of axioms. 
For example, induction over the natural numbers needs to be explicitly provided as an axiom as shown in section~\ref{sec:weblf:taste}.
More modern proof assistants have founds solutions to this limitation by providing the programmer with a structured way of defining inductive types. In the next section we show how these inductive types work in an extension to WebLF called WebPie. Subsequently, we show how their formalisation and how to implement them.

\section{WebPie: Adding Inductive Types and Case Matching}
Instead of letting the programmer define separate axioms one-by-one in WebPie the programmer can define a set of related axioms in one go. 

Consider the example of natural numbers, in WebLF the programmer needs to define three axioms one for introducing the \lstinline{Nat} set and two for the constructors.
In WebPie these three related axioms can be defined as one inductive type, shown in Figure~\ref{webpie:nat}. 

\begin{figure}[thp] 
\begin{lstlisting}[inputencoding=utf8]
Inductive Nat : Set :=
  | Zero : Nat 
  | Succ : Nat  → Nat;
\end{lstlisting}    
\caption{WebPie: Inductive definition of natural numbers.}
\label{webpie:nat}
\end{figure} 

At first sight the programmer has not gained a lot by using a slightly different syntax. 
It is however, important to note that the programmer implicitly indicates that the definition of the set \lstinline{Nat} is closed. Once the inductive type is defined there is no mechanism in place to extend it.
Because inductive definitions are closed the only possible constructors for creating elements in \lstinline{Nat} are \lstinline{Zero} and \lstinline{Succ}. 
Armed with this knowledge it becomes feasible to provide better support to the programmer.
For example, it becomes possible to provide case matching and to
actually prove the induction principle instead of providing it as an
axiom. Without demanding that \lstinline{Nat} is closed, the
programmer could extend the constructors of the \lstinline{Nat} set
which would make type checking much more difficult.

\subsubsection*{Case Matching}
Case matching allows the programmer to inspect a value of an inductive type and determine with which constructor the value was created. 
Important to realise is that for all of the different cases the case match should return a value of the same type. 
As an example of case matching consider the implementation of addition in Figure~\ref{webpie:nat_case}.

\begin{figure}[thp] 
\begin{lstlisting}[inputencoding=utf8]
def add(x:Nat,y:Nat) : Nat {
  <(λn:Nat.Nat)> 
  match x with {
      Zero   => y;
      Succ   => (λn:Nat.(Succ (add n y)))
  }   
};
\end{lstlisting}    
\caption{WebPie: Case matching and recursion.}
\label{webpie:nat_case}
\end{figure} 

In this example case match \lstinline{<λn:Nat.Nat>} indicates that we case match over an element in \lstinline{Nat} and that we will return a \lstinline{Nat}. \lstinline{ match x with} indicates that we case match over the variable \lstinline{x}. There are two constructors to consider: \lstinline{Zero} and \lstinline{Succ}. When \lstinline{x = Zero} we simply return \lstinline{y}. In case the natural number was constructed with \lstinline{Succ} for example \lstinline{(Succ Zero)}  the function \lstinline{(λn:Nat. (Succ (add n y)))} will be applied to the arguments of the constructors, i.e. \lstinline{Zero}. 
The function returns a new number by adding its argument \lstinline{n} to \lstinline{y} and subsequently wraps it into a \lstinline{Succ}.
Note that from an abstract point of view case matching is very much like function application, you supply it with an argument and it returns a certain value. 

\subsubsection*{Dependent case matching}
Sometimes regular case matching is not powerful enough to capture the dependencies in our applications. 
In those cases we can make use of dependent case matching to make an explicit dependency between the return type of a case match and the value over which we case match. 
As an example, consider the proof of the induction principle over the natural numbers shown in Figure~\ref{fig:webpie:indnat}. 

\begin{figure}[h!] 
\begin{lstlisting}[inputencoding=utf8]
def nat_ind
( P  : Nat → Set,
  f  : (P Zero),
  fn : (Πn:Nat.(P n) → (P (Succ n))),
   n : Nat 
) :  (P n)  
{
   <(λn:Nat.(P n))> 
   match n  with {
      Zero  => f;
      Succ  => (λn:Nat. (fn n (nat_ind P f fn n)))
   }   
};
\end{lstlisting}    
\caption{WebPie: Induction principle over natural numbers.}
\label{fig:webpie:indnat}
\end{figure}

The encoding of the induction principle at the type level is the same as in section~\ref{sec:weblf:taste}.
The big difference here is that we do not state the induction principle as an axiom. 
The body of the function provides a proof that the reasoning principle is indeed correct. 

Let us look at the proof term in detail. First of all, it is a lambda abstraction which expects one argument of type \lstinline{Nat}, given this argument it produces a proof \lstinline{(P n)} by case matching over n. 
Because we need to produce a different proof given the argument $n$ we make use of dependent case matching. 
In the example, \lstinline{<λn:Nat.(P n)>} indicates that we case match over an element in \lstinline{nat} and that all the cases of the case match will return a proof \lstinline{(P n)}. 

Let us look into detail what this entails when \lstinline{n} equals \lstinline{Zero}. In this case the return type can be computed by substituting \lstinline{n} for \lstinline{Zero} in the body of the \lstinline{λn:Nat.(P n)} , which equals \lstinline{(P Zero)}. The implementation for the \lstinline{Zero} case is easy because the argument \lstinline{f} has exactly type \lstinline{(P Zero)}. 

In the second case the return type of the match should be \lstinline{(P n)}. Implementing this case is a bit more difficult but we conveniently receive a function \lstinline{fn} which when we provide a proof of \lstinline{(P x)} will return us a proof of  \lstinline{(P (Succ x))} which is equal to \lstinline{(P n)}. Obtaining the proof of \lstinline{(P x)} is done by a recursively call to \lstinline{nat_ind}. 

Readers who are worried at this point about the last recursive call, can be assured that this recursion will stop eventually because the last argument \lstinline{x} decreases in each recursive call.
As this argument decreases in every recursive call, eventually the base case (\lstinline{n} equals \lstinline{Zero}) will be reached. In section \ref{guard}, we explain how to the implementation deals with checking this (guard) condition.

\subsubsection*{Equality over natural numbers}
We still need one ingredient to start making simple proofs over \lstinline{Nat}, namely equality. 
There are many ways to encode equality in a proof assistant but here we define equality very specifically for natural numbers by defining an inductive type \lstinline{Eq} (Figure~\ref{fig:webpie:eq}). It has two constructors, \lstinline{Eq_Rfl} and \lstinline{Eq_Succ}. The first constructor encodes that all natural numbers are equal to itself. The second constructor encodes that when we have a proof that two numbers are equal to each other we can conclude that adding one to each of the numbers results in equal numbers. 

\begin{figure}[h!] 
\begin{lstlisting}[inputencoding=utf8]
Inductive Eq : Nat → Nat → Set := 
 | Eq_Rfl  : Πn:Nat.(Eq n n)
 | Eq_Succ : Πx:Nat.Πy:Nat.
             (Eq x y) → 
             (Eq  (Succ x) (Succ y));
\end{lstlisting}    
\caption{WebPie: Equality over numbers.}
\label{fig:webpie:eq}
\end{figure}

Having defined natural numbers, addition, the induction principle over natural numbers and equality of natural numbers it becomes possible to make some small proofs. A first proof \lstinline{add_zero} shows that for any number \lstinline{x} adding \lstinline{Zero} to the left of that number equals \lstinline{x}. 
The proof is constructed by applying the constructor \lstinline{Eq_Rfl} which returns us a proof \lstinline{(Eq x x)}. WebPie normalises the \lstinline{(add Zero x)} to \lstinline{x} and so the term is well typed.

A second proof \lstinline{add_x_zero} shows that for any \lstinline{Nat} adding \lstinline{Zero} to the right equals that \lstinline{Nat}. 
This proof is a bit more difficult because WebPie cannot normalise \lstinline{(add x Zero)} to \lstinline{x}.  The reasons why WebPie cannot normalise this is because the function \lstinline{add} case matches over the first argument, which could be any \lstinline{Nat}. 
We can however, make use of our induction principle \lstinline{nat_ind} defined earlier. 

The function \lstinline{nat_ind} expects as first argument a function which given a \lstinline{Nat} gives us a proposition over this \lstinline{Nat}. In our case we want to prove the proposition \lstinline{(Eq (add n Zero)  n)}.  

The second argument is a proof for the case where the \lstinline{Nat} is \lstinline{Zero}. Substituting \lstinline{n} for \lstinline{Zero} in our proposition gives us \lstinline{(Eq (add Zero Zero)   Zero)}. We can provide a proof for this proposition by applying \lstinline{Eq_Rfl} to \lstinline{Zero}. This gives us a proof that \lstinline{(Eq Zero Zero)}. This is not exactly the same as \lstinline{(Eq (add Zero Zero)  Zero)} but by normalisation WebPie deduces that \lstinline{(add Zero Zero)} equals to \lstinline{Zero} and hence that the two proofs are equal. 

The third argument is a function which when given a proof for \lstinline{(P n)} returns us a proof for \lstinline{(P (Succ n))}. The proof is created by making a new lambda abstraction which takes these arguments and produces the proof \lstinline{(P (Succ n))} by applying \lstinline{Eq_Succ}. 
Note that \lstinline{Eq_Succ} will return a proof of \lstinline{(Eq  (Succ (add n Zero))  (Succ n))}, this is not syntactically equal to \lstinline{(P (Succ n))} which by simple substitution equals to \lstinline{(Eq  (add (Succ n)  Zero) (Succ n))}. The trick here is that WebPie normalise the subterm \lstinline{(add (Succ n)  Zero) } to \lstinline{(Succ (add n Zero))} making the two terms syntactically equal.

\begin{figure}[thp] 
\begin{lstlisting}[inputencoding=utf8]
def add_zero(x:Nat) : (Eq (add Zero x) x) {
   (Eq_Rfl x)
};

def add_x_zero(x:Nat): (Eq (add x Zero) x) {
	(nat_ind 
	 (λn:Nat.(Eq (add n Zero) n))
	 (Eq_Rfl Zero)
	 (λn:Nat.(λIH:(Eq (add n Zero) n).
		  (Eq_Succ (add n Zero) n IH)))
	 x)
} 
\end{lstlisting}    
\caption{WebPie: Example proofs over natural numbers.}
\end{figure}

\section{WebPie Formally}

We define WebPie as an extension to WebLF, the formal exposition here is a simplification of the exposition as shown in \cite{eduardo94}. 
The extension to the syntax of WebLF are shown in Figure~\ref{fig:webpie:syntax}.
The extensions are inductive definitions ${\tt Ind}$, constructors ${\tt  Constr}$, case matching ${\tt Match}$, and recursive functions ${\tt Fix}$ . 
Inductive definitions are defined by specifying its name $x$ and type $e_t$ together with a list of constructor names $x_i$ and their respective types $e_i$.
A constructor is defined as a pair of the inductive type $e_I$ and a number $n$ indicating the index of the constructor in the inductive type. 
A dependent case match expects the type of the case match $e_{ct}$ , the expression over which we case match and a list of deconstructors. 
Finally, recursive functions are defined by specifying a name $x$ and the type of the recursive function $e_t$. 
Note that in the formalisation we also need to specify an index $k$. This index is used to help the type-checker in determining whether executing the recursive function will stop.   
In essence it indicates which argument of the recursive function is becoming smaller with each recursive call. 
The intuition is that if the type-checker can determine that an (inductive) argument of a  recursive function becomes smaller in each recursive call then eventually the function needs to stop.  
\begin{figure}[h!]
\center
  \begin{tabular}{ r l l r }
  	$e $ & $::=$ & ... & $$\\ 
	 & $|$ & ${\tt Ind}(x:e_t)\{ x_1:e_1 | ... | x_2:e_n \}$  & $Ind.~Type$\\	 	
	 & $|$ & ${\tt Constr}(e_I,n) $  & $Constructor$\\
	 & $|$ & ${\tt Match}(e_{ct},e_m) \{ x_{c1} \Rightarrow e_1 | ... | x_{cn} \Rightarrow e_n \} $  & $Case~Match$\\
	 & $|$ & ${\tt Fix_k}(x:e_t)\{e_b \} $  & $Fix$\\
 \end{tabular}
  \caption{WebPie Syntax Extensions.}
  \label{fig:webpie:syntax}
\end{figure} 

\subsection{Type Checking Inductive Types}
Type checking of inductive types is governed by three typing rules,
one for defining the inductive type, one for type checking
constructors and finally one for case matching.

\subsubsection*{Inductive definitions}
In order to typecheck the definition of an inductive type we need to verify two things. 
First, all the constructors of the inductive type should return an object in the same universe. 
Which should also be the same universe as indicated in the type annotation when defining the inductive type. 
For example, in Figure~\ref{webpie:nat} the type annotation indicates that objects of type \lstinline{Nat} should live in universe \lstinline{Set}. We should verify that under the assumption that \lstinline{Nat:Set} all of the constructors also return something in type universe \lstinline{Set}.

In order to get the universe of a $\Pi$ type we define a function $\Upsilon$ which recursively traverses the type until it encounters a universe. 

\begin{figure}[h]
\center
\begin{tabular}{ r c r }
	$ \Upsilon( \Pi x:e_m.e) $ & $::=$ & $\Upsilon(e) $\\ 
	$ \Upsilon(Type_i) $ & $::= $ & $  Type_i $\\ 
\end{tabular}	
\end{figure}	

The second property we need to verify in order to typecheck an inductive definition is that the return type of all the constructors is an element of the inductive type which is being defined.
It would of course be a type error when the constructor of a \lstinline{Nat} returns a \lstinline{Bool}.
The set of all possible well formed constructors for an inductive type with name $X$ is generated by the syntax $Co$. 
With the restriction that $X$ does not occurs free in $e_n$ and $e_m$.
The intuition behind this syntax is that the type should end in an application of the inductive type being defined. We use shorthand notation $C(e,x)$ to indicate that the type represented by $e$ is a valid constructor for the type with name $x$.

\begin{figure}[h]
\center
\begin{tabular}{ r c r }
	$ Co  $ & $::=$ &  $(X ~\overline{e_n})~|~P \rightarrow Co~|~\Pi x:e_m . Co $\\ 
\end{tabular}	
\end{figure}

After forming an intuition about the two properties that need to be verified the typing rule for inductive definitions (\textsc{T-Ind}) follows directly as shown in Figure~\ref{fig:webpie:typing}. \\

\subsubsection*{Case Matching}

The type checking rule for a match expression is complicated by the fact that there are multiple cases.
Consider again the case match for the successor case in Figure~\ref{fig:webpie:indnat}. 
\begin{lstlisting}
 <(λn:Nat.(P n))> 
   match n  with {
      Zero => f;
      Succ => (λx:Nat.(fn x (nat_ind P f fn x)))
   }
\end{lstlisting}
In order to typecheck this expression the type checker first needs to ensure that the type of \lstinline{n} is an inductive type. Once the type checker is sure that \lstinline{n} is an inductive type it needs to ensure that:

\begin{itemize}
\item The list of cases provided by the programmer corresponds with the inductive definition. 
\item The return type of all the cases corresponds to the type annotation \lstinline{(Πn:Nat.(P n))} instantiated with the particular case.  
\end{itemize}

The rule {\textsc T-Match} encodes these two guards as shown in Figure~\ref{fig:webpie:typing}. 
Checking that the list of cases is correct, is done by verifying that the list of constructor names in the cases corresponds to the definition i.e. $x_{c1} ... x_{cn}$. 
Note that the typing rule enforces the programmer to write the cases in the same order as in the definition. 
It would not be that difficult to allow for permutations.  

Checking that the type of all the cases is correct is a bit more difficult. 
In the example, there are two cases \lstinline{Zero} and \lstinline{(Succ x)}. 
To check that the right hand side of the case match is correct we first need to compute what the expected type for that case match should be. 

This is done by apply the matched term to $e_{ct}$ \lstinline{(λn:Nat.(P n))} we respectively get \lstinline {(P Zero)} and  \lstinline {(P (Succ x))}. In the typing rule this is ensured by a meta predicate $S$. This meta predicate given the type of the constructor, the matching type and the constructor builds the appropriate type. 

\begin{align*}
	S(\Pi x:e_{tx}.e_r ,e_{ct},c) & = \Pi x:e_{tx}.S(e_r,e_{ct},(c~x))    \\
	S((x_i~\overline{e_a}),e_{ct},c ) &  =  (e_{ct}~\overline{e_a}~c)
\end{align*}

%

Finally, if all the cases are valid then the type of the entire case match can be derived by applying the type $e_t$ to  the parametric arguments of the inductive type over which we case match $\overline{x_m}$ together with the expression over which we case match $e_m$.    

\subsubsection*{Constructors}
Lastly, constructors are represented by an index $i$ and an inductive definition $I$. 
Typechecking a constructor is done by verifying that the inductive definition is well typed and ensuring that the index is within bounds of the inductive definition (\textsc{T-Constr}). 

\begin{figure}[h!] 
\createbox{ $\Gamma \vdash  x : $T}
\begin{mathpar}
   \inferrule*[Right=(T-Ind)]
  { \Upsilon(e_t) = u \\ 
    \and \forall i = 1 .. n~
    \Gamma,x:e_t \vdash e_i : u \and 
     C(e_i,x) }
  { \Gamma \vdash {\tt Ind}(x:e_t)\{ x_1:e_1 | ... | x_2:e_n\} : e_t }

  \inferrule*[Right=(T-Constr)]
  { 1 < i <= n
  \\ 
   I = {\tt Ind}(x:e_t)\{ x_1:e_1 | ... | x_2:e_n\}  \and 
    \Gamma \vdash I  : e_t  
     }
  { \Gamma \vdash  {\tt Constr}(i,I) :  [ x \mapsto I ]  e_i  }

  \inferrule*[Right=(T-Match)]
  {    I = {\tt Ind}(x:e_t)\{ x_{c1}:e_{t1} | ... | x_{cn}:e_{tn}\}\\
      \Gamma \vdash  e_{ct} : \Pi\overline{x_p:e_{pt}}.  (I~\overline{x_p}) \rightarrow  {\tt Type_i}  \\
      \Gamma \vdash  e_m : (I ~ \overline{x_m}) \\ 
      \forall i = 1..n, \Gamma  \vdash  e_i : S(e_{ti},e_t,x_{ci})
   }
  { 
  \Gamma \vdash  {\tt Match}(e_{ct},e_m)  \{ x_{c1} \Rightarrow e_1 | ... | x_{cn} \Rightarrow e_n \} : 
  	(e_t~\overline{x_m}~e_m)
  }   
  
  \inferrule*[Right=(T-Fix)]
  {   
      \Gamma \vdash  e_t :  {\tt Type_i}  \\
      \Gamma, x:e_t \vdash  e_b : e_t \\
      \beta\{f,k,arg_k(e_b),\emptyset,e_b\}
   }
  { 
  \Gamma \vdash {\tt Fix_k}(x:e_t)\{e_b \}  :  e_t
  }

\end{mathpar}
  \caption{WebPie Typing Extensions.}
  \label{fig:webpie:typing}
\end{figure}

\subsection{Type Checking Recursive Functions}
\label{guard}

Because a dependently typed language encodes a logic, care needs to be taken in order to make sure that the programmer can only define pure terminating total functions. While some dependently typed programming languages allow the programmer to omit these totality checks~\cite{JFP:9060502} it is important to know that in such case the programming language does not ensure sound theorem proving. 

Making sure that programs terminate is a well known problem in computer science for which no terminating algorithm exists.
Therefore, any termination checking function is a conservative approximation of the set of all functions which terminate. 
When the termination check succeeds the programmer is sure that the function always terminates. 
On the other hand when the termination check fails the function might still terminate in practice. 
Even full fledged proof assistants such as Coq disallow the definition of certain obviously terminating functions. 
In this section we explain the basics of a simple termination checker. 
The basic idea of the termination checker is to make sure that at least one of the arguments of all recursive calls has an inductive argument which is decreasing. 
The guard condition is enforced by the meta predicate $\beta$. 

\begin{align*}
	\beta\{ f, k, x_k,x_g, e \}  & =   true & f \notin fv(e)    \\
	\beta\{ f, k, x_k,x_g, x_n \}  & =  x_n \neq f  \\
	\beta\{ f, k, x_k,x_g,   {\tt Match}(e_{ct},e_m)  \{ x_{ci} \Rightarrow e_i \}   \}  & =     
	\beta\{ f, k, x_k,x_g \cup args(e_i),  body(e_i)\} \wedge
	&  e_m \in  x_g  \\
	& ~~~~\beta\{ f, k, x_k,x_g, e_t\}  \wedge \beta\{ f, k, x_k,x_g, e_m\}  \\
	\beta\{ f, k, x_k,x_g, (x_n~\overline{x_p}) \}   & =  \forall x_p \in e . \beta\{f,k,x_k,x_g,x_p\} ~ \wedge~ \overline{x_p}_k \in x_g 
	\cup x_k    & x_n = f \\
	\beta\{ f, k, x_k,x_g, e \}  & =  \forall e_s \subset e . \beta\{ f, k, x_k,x_g, e_s \}  \\
\end{align*}

The basic idea of this meta predicate is to verify whether the k\textsuperscript{th} argument of each recursive call is guarded by a deconstruction. 
At the same time the meta predicate needs to make sure that aliases to  the recursive function do not leak. 
In this case we simply do not allow the function to be passed on to other functions or to be returned in any form. 

The meta predicate $\beta$ first verifies whether a reference to the recursive function $f$ is present in the expression $e$, if this is not the case all the recursive calls (none) will terminate. 
When we encounter a variable that is ok as long as it is not the recursive function $f$ which is being defined. 
The reason we need to exclude all references to the recursive function is because otherwise we would need to include some form of alias tracking which would significantly increase the complexity of the termination check. 

The two most interesting cases are the ones for a case match and application.  
When encountering  a case match over a variable in our guarded variables we recursively apply $\beta$ over the deconstructors.
For each of these deconstructors the arguments of the deconstructors are added to the set of guarded variables.  
Because the recursive function can be used at the type level we also need to verify whether any recursive call in the type $e_t$ is also guarded. 

When verifying the guard condition over an application $ (x_n~\overline{x_p})$ where $x_n$ is the recursive function $f$ being defined the  k\textsuperscript{th}  argument needs to be guarded and all the arguments need to pass the guard condition. 
Finally, for all other case the guard condition is recursively applied to all the subexpressions. 

\subsection{Normalisation}

Next to extending the typing rules we also need to extend
the normalisation rules to account for case matching. When
case matching the term we first normalise the expression on
which we case match. If it normalises to a constructor we
can further normalise by substituting the arguments to the
constructor in the right hand side of the corresponding case
match E-Match1. Otherwise the whole term normalises to a
new match where $e_m$ has been replaced by its normal form
$e'_m$.

 \begin{figure}[h!] 
\createbox{ $e \Downarrow n $}
\begin{mathpar}

   \inferrule*[Right=E-Match-1]
  {  e_m \Downarrow ({\tt Constr}(i,I) \overline{e_a}) \\
    [ \overline{x_{ai}} \mapsto \overline{e_{ai}})] \Downarrow e'_i 
    }
  {   {\tt Match}(e_{ct},e_m) \{ (x_{c1}~\overline{x_{a1}})\Rightarrow e_1 | ... | (x_{cn}~\overline{x_{an}}) \Rightarrow e_n \}  
      \Downarrow  
        e'_i   
  }

   \inferrule*[Right=E-Match-2]
  {  e_m \Downarrow e'_m }
  {   {\tt Match}(e_{ct},e_m) \\ \{ (x_{c1}~\overline{x_{a1}})\Rightarrow e_1 | ... | (x_{cn}~\overline{x_{an}}) \Rightarrow e_n \}  
     \Downarrow     \\\\
       {\tt Match}(e_{ct},e'_m) \\ \{ (x_{c1}~\overline{x_{a1}})\Rightarrow e_1 | ... | (x_{cn}~\overline{x_{an}}) \Rightarrow e_n \}  
  } 

\end{mathpar}
  \caption{WebPie Normalisation Extensions.}
  \label{fig:webpie:Normalisation}
\end{figure}

\vspace{-0.8em}
\section{Implementing WebPie}
The implementation of inductive types largely follow the
extensions show in the formalisation. We first extend the
union for Expressions with inductive definitions, constructors and case matching. 
Because the JavaScript extension for ADT’s does not have native support for arrays the type of
constructors and destructor is set to be * which means they could be anything. 
In our implementation constructors is a list of JSON objects with two fields \lstinline{name} and \lstinline{ctype}. 
Destructors are represented as an array of JSON objects with two fields \lstinline{name} and \lstinline{expr}.

\begin{figure}[h]
\begin{lstlisting}[keywords={union,deriving,function,case,return}]{JavaScript}
union Expression {
   // .. extending WebLF
   Inductive { name         : String, 
               arity        : Expression,
               constructors : * },
   Constr    { index        : Number,
               inductive    : Expression},             
   Match     { carrier      : Expression,
               expr         : Expression, 
               destructors  : * }            
} deriving (adt.Eq,adt.Extractor,adt.ToString)              
\end{lstlisting}
\caption{WebPie: Expression extension.}
\end{figure}

\begin{figure}[h]
\begin{lstlisting}
function check_inductive(n,a,cst,ctxt) {
	var u = getUniverse(a);
	cst.forEach(function(cst) {
	   var tCst = type_check(extend_type(n,a,ctxt),cst[1]);
	   var eq   = equal_expr(tCst,u);
	   positive(cst[1],n) );
	}); 
	return a;
}  
\end{lstlisting}
\caption{Type Checking Inductive Definitions.}
\label{fig:webpie:type_inductivedef}

\end{figure}

\begin{figure}[h!] 
\begin{lstlisting}
 function check_match(c,e,cases,ctxt) {
      var IndT    = normalise(type_check(ctxt,e),ctxt);
      var c_type  = normalise(type_check(ctxt,c),ctxt);
      var Ind     = findInductive(IndT); 
      var pargs   = countArgs(Ind.arity);
      verify_c_type(c_type,Ind,pargs);
      var constructors = Ind.constructors;
      checkMatchLength(constructors,cases);
      cases.map( function(cs,i) { 
           var cstT = constructors[i].ctype;
           var case_type  = normalise(makeTypeForCase(cstT,c,cs.name),ctxt);
           var e_it = type_check(ctxt,cs.expr); 
           var x = check_equal(e_it,case_type,ctxt);    
      }); 
      return normalise(App(makeRApp(c,IndT,pargs),e),ctxt); 
 }     
\end{lstlisting}
\caption{Type Checking a Match Expression.}
\label{fig:webpie:type_match}
\end{figure}

Extensions to the implementation of the type checker
consists of adding cases for the constructors, inductive types
and case matching. Because the programmer cannot type in
a constructor himself typing the constructor is as simple as
returning the field of the inductive definition.

Typing inductive definitions consist of checking that the
universes correspond and that the constructors are well
formed, as shown in Figure~\ref{fig:webpie:type_inductivedef}. 
In our implementation getUniverse and positive corresponds to $\Upsilon$ and $C(x, e)$ in the formalisation.

Finally type checking case matches (Figure~\ref{fig:webpie:type_match}) differs a little bit from
the formalisation because the syntax of the formalisation and
the implement differ slightly. In the implementation the
programmer needs to supply an abstraction with the
correct number of arguments for the constructor i.e. 
\lstinline{Succ => (λn:Nat. (Succ (add n y))) }instead of 
\lstinline{(Succ n)=> (Succ (add n y)))}.
The code first verifies whether the value over which
we case match is an inductive type. Then it verifies whether
the constructors match the cases. We then verify for each of
the constructors that the right hand side of the cases match.
In the implementation we do this by explicitly building up a
$\Pi$ type and apply normalisation with \lstinline{makeTypeForCase}.

\section{Safe Printf in WebPie}
As a final example of using WebPie, we show how to encode the safe \lstinline{printf} example of the introduction. 
As WebPie does not have any primitives for input output we merely simulate a safe \lstinline{printf}.

The implementation of our safe \lstinline{printf} (shown in Figure~\ref{fig:webpie:printf}) starts by defining an inductive type \lstinline{FormatString} to construct format strings. 
In our case the format string is either \lstinline{End} or a type appended to the format string, i.e. a list of types. 
Given a list of types the function \lstinline{computeType} gives us the type of a function that expects all the argument in the list and eventually returns the type \lstinline{Void}. 
Finally, the function \lstinline{printF} is a dependently typed function that given a format string \lstinline{e} has the type \lstinline{(computeType e)}.
The implementation of \lstinline{printF} makes use of dependent pattern matching on the format string. If the end of the format string has been reached it will simply return \lstinline{Null}. 
In the other case it will create a new function that expects the head for the format string and recursively creates a new function based on the rest of the format string. 

\begin{figure}[h!] 
\begin{lstlisting}
Inductive FormatString : Type :=
  | End  : FormatString
  | Cons : Set -> FormatString -> FormatString;

def computeType(e:FormatString) : Set  {
  <(Πf:FormatString.Set)> 
  match e with  {
     End   => Void ;
     Cons  => (λhf:Set.
               (λrf:FormatString.
                  hf -> (computeType rf)))
  }
};

def printF( e:FormatString ) : (computeType e) {
  <(Πf:FormatString.(computeType f))> 
  match e with  {
     End => Null;
     Cons => (λhf:Set.
               (λrf:FormatString.
                 (λx:hf.
                  (printF rf))))
  }
};
\end{lstlisting}
\caption{Implementation of printF.}
\label{fig:webpie:printf}
\end{figure}


\vspace{-2.5em}
\section{Conclusion}
It is our hope that this hands-on-approach towards the implementation of a small dependently typed programming language is a good addition to the more formal expositions of dependent typing. We gave a detailed explanation of the most important concepts of dependent typing and have illustrated with examples how these concepts can be used. We have avoided concerns such as efficiency and error handling to focus on the essence of the language.  
With this introduction to WebPie the reader can obtain a good understanding of how inductive datatypes, dependent case matching and guarded recursive functions work in dependently typed programming languages and can start reasoning about extending the language.

\bibliographystyle{eptcs}
\bibliography{cites}

\newpage 

\appendix

\section{WebLF Rules for Normalisation}
 \begin{figure}[h!] 
\createbox{ $e \Downarrow n $}
\begin{mathpar}
%
 \inferrule*[Right=(N-App-Lam)]
  { e_f \Downarrow \lambda x:e_t . e_b \\  e_a \Downarrow  e_{a'}  \\  [x\mapsto e_{a'}] e_b \Downarrow n}
  { e_f~e_a \Downarrow n} 
 \\ 
  \inferrule*[Right=(N-App-e)]
  { e_f \Downarrow e_f' \\  e_a \Downarrow  e_{a'} }  
  { e_f~e_a \Downarrow e_f'~e_a'} 
   
                        
 \inferrule*[Right=(N-Abs)]
  {  e_t \Downarrow n_t \\ e_b \Downarrow n_b }
  { \lambda x:e_t . e_b \Downarrow \lambda x:n_t . n_b} 
 \hva \\ \and   
   \inferrule*[Right=(N-Pi)]
  {  e_t \Downarrow n_t \\ e_b \Downarrow n_b }
  { \Pi x:e_t . e_b \Downarrow \Pi x:n_t . n_b} 

\end{mathpar}
  \caption{WebLF normalisation rules.}
  \label{fig:Normalisation}
\end{figure}

\section{Using Webpie}
In this appendix we showcase some simple theorem proving and programming examples in Webpie.
Concretely we show how to encode the first examples of the software foundations book volume 1~\cite{Pierce:SF1}. 

\subsection{Data and Functions} 
In the example below we define a new inductive Set, the name of the new set is \lstinline{day} and it has seven constructors, one for each day of the week.

\begin{figure}[h!] 
 
\begin{lstlisting}
Inductive day : Set :=
  | monday     : day 
  | tuesday    : day 
  | wednesday  : day 
  | thursday   : day 
  | friday     : day 
  | saturday   : day 
  | sunday     : day;
\end{lstlisting}

\caption{Inductive definition of day.}
\end{figure} 

Once an inductive set has been defined we can write functions which take elements of \lstinline{day} as arguments or return elements of the set \lstinline{day}. 
For example, we can define a function which given a day returns the next weekday.

In the definition below we indicate that we will define a new function, the name of this function is \lstinline{next_weekday} it takes one argument \lstinline{d} which needs to be an element of day and we return an element of type \lstinline{day}.

In order to inspect which particular day the function was called with, we make use of case matching. 
When matching over an inductively defined set the programmer needs to specify the type over which will be case matched and what the return type of each of the cases such be. 
In our case we match over a day and return a new day, in the code this is indicated by the annotation \lstinline{<λx:day.day>}

\begin{figure}[h!] 
\begin{lstlisting}
def next_weekday(d:day) : day {
  <λx:day.day> 
  match d with {
   monday    => tuesday;
   tuesday   => wednesday;
   wednesday => thursday;
   thursday  => friday;
   friday    => monday;
   saturday  => monday;
   sunday    => monday
  }   
};
\end{lstlisting}
\caption{A function over the day type.}
\label{fig:webpie:next_weekday_function}
\end{figure} 
  
Next to inductively defined sets we can also define propositions. 
While it is tempting to think about propositions as things which are true this is certainly not the case.
There are many propositions that can be stated but much fewer which can actually be proven to be true.

We start with a well known proposition namely equality. 
What does it mean for two things to be the same?
One way of defining that two things are same is to say that two things
are the same if they are exactly the same thing. Each object is the same as itself.
To capture this idea we define an inductively defined relation \lstinline{eq} which has only one constructor  \lstinline{eq_refl}.
In order to avoid having to define equality propositions for each and every type we define we can make the equality proposition parametric in the type over which we define the equality. The type annotation \lstinline{ΠT:Set.T->T->Set} indicates that the equality type expects a type \lstinline{T} and two values of type T. 
  
\begin{figure}[h!] 
 
\begin{lstlisting}
Inductive eq: ΠT:Set.T->T->Set :=
 | eq_refl : ΠT:Set.Πx:T.(eq T x x); 
\end{lstlisting}
\caption{Parametric Equality Proposition.}
\label{fig:webpie:eq_prop}
\end{figure}

\subsection{Proof by normalisation} 
 
Armed with our equality proposition we can now start testing whether our  definition of \lstinline{next_weekday} is doing what we suspect it is doing.  
For example, we can now write a proof to show \lstinline{(eq day (next_weekday monday) tuesday)}.  Our hopes are that when we evaluate \lstinline{(next_weekday monday}) it will give us \lstinline{tuesday}. 
So in order to prove \lstinline{(eq day (next_weekday monday) tuesday)},  we need to construct a proof \lstinline{(eq day tuesday tuesday)}. 
 Constructing such a proof is not that hard, we use the only available constructor for eq namely \lstinline{eq_refl}, the first argument it expects is a set, and as we are  trying to construct a proof over days we supply it with the type \lstinline{day} as the first argument. 
 Subsequently, we need to supply it with an element of day which is in our case is  \lstinline{tuesday}.
 The resulting term has type \lstinline{(eq day tuesday tuesday)}. The typechecker will normalise \lstinline{ (next_weekday monday)}  to  \lstinline{tuesday}  during type checking and can conclude that what we wanted to prove is indeed correct. 
\begin{figure}[h!] 
 
\begin{lstlisting}
def test_next_weekday(x : Void) : (eq (next_weekday monday) tuesday) {
	(eq_refl tuesday)
}
\end{lstlisting}
\caption{Testing next weekday.}
\label{fig:webpie:next_weekday_proof}
\end{figure} 

\subsection{Proof by rewriting}
We have now shown how to use normalisation to prove some easy propositions. 
Unfortunately, we need some stronger proof techniques when we want to prove some more interesting propositions. 
As an example, imagine that we know that some number $n$ is equal to some other number $m$. 
Knowing that they are equal we should be able to prove that  $n+n = m +m$. 
 
In theorem provers such as coq the programmer has the ability to simply rewrite the goals in case she has a proof that two terms are equal. 
In that case if the programmer has a proof that $n=m$ she can rewrite $n+n= m+m$ to $n+n = n+n$ and then use reflexivity. 

If we want to be able to use such a  rewriting technique in WebPie we first need to prove rewriting. 
We define a new function \lstinline{rewrite} which given two elements \lstinline{x,y} of type \lstinline{T}, a proof \lstinline{p} that the two elements are equal and a proof of a proposition \lstinline{(P x)} gives us a proof of \lstinline{(P y)}. Note that \lstinline{P} itself is a function which returns a proposition when given a value with type T.  

The actual proof for rewrite is the body of the function. 
In order to fully understand how this proof works it is recommended to keep the {\sc T-Match} rule at hand. 
In essence the case match type indicates that when case matching over a proof \lstinline{(eq T p q)} it will return a proof \lstinline{(P q)}. 
The case match type \lstinline{ΠT:Set.Πp:T.Πq:T.(eq T p q) -> (P q)} looks a bit more complicated because the equality type is parametric. 
To account for this the dependent case match needs to make these parameters of the equality type explicit, namely a set \lstinline{T}, and two elements of that set. As the equality type only has one constructor (\lstinline{eq_refl}) there is only one case to consider. 
This constructor expects two arguments, a type \lstinline{T} and an element of that type. 
In the \lstinline{eq_refl} case we can conveniently return \lstinline{H} which has type \lstinline{(P x)}.

The attentive reader might be confused at this point because the proof \lstinline{H} is clearly not a proof of \lstinline{(P y)}.
The expected type of the \lstinline{eq_refl}  case is however constructed with the $S$ helper function:\\
\lstinline{ΠT:Set.Πx:T.((λT:Set.λp:T.λq:T.(eq T p q) -> (P q) ) T x x (eq T x x))} which after normalisation reduces to: \lstinline{ΠT:Set.Πx:T.(P x)}.
Hence the proof \lstinline{(P x)} is exactly what we needed in this case. Finally, because this is the only way we can construct a eq type we are finished. 
  
\begin{figure}[h!] 
\begin{lstlisting}
def rewrite(T:Set,P:(T->Prop),x:T,y:T,p:(eq T x y),H:(P x)) : (P y) {
  <(λT:Set.(λp:T.(λq:T.(λeq:(eq T p q). (P q)))))>
  match p with {
   eq_refl => (λT:Set.(λx:T.H))
  }
};
\end{lstlisting}
\caption{Proving rewrite.}
\label{fig:webpie:rewrite}
\end{figure} 

Now that we have a proof of rewrite we can use it to prove our proposition that when two numbers are equal to each other adding those numbers together also result in equal numbers as shown in Figure~\ref{fig:webpie:use_rewrite}. 

\begin{figure}[h!] 
 
\begin{lstlisting}  
def plus_id_example(n:nat,m:nat,h1:(eq nat n m)) : 
(eq nat (plus n n) (plus m m)) {
(rewrite nat 
         (λx:(nat). (eq nat (plus n n) (plus x x))) 
         n m h1 
         (eq_refl nat  (plus n n)))
};
\end{lstlisting}
\caption{Using the rewrite function}
\label{fig:webpie:use_rewrite}
\end{figure}   
  
\newpage

\newpage

\section{Proofs over natural numbers}

\begin{lstlisting}
Inductive Nat : Set :=
  | Zero : Nat 
  | Succ : Nat  -> Nat;
Inductive Plus : Nat -> Nat -> Nat -> Set :=
  | Plus_Zero  : Πn:Nat.(Plus Zero n n)
  | Plus_Succ  : Πx:Nat.Πy:Nat.Πs:Nat.Πp:(Plus x y s).
                          (Plus (Succ x)  y (Succ s));
                
Inductive Eq : Nat -> Nat -> (Type 1) :=  
 | Eq_Rfl  : Πn:Nat.(Eq n n)
 | Eq_Succ : Πx:Nat.Πy:Nat.(Eq x y) -> (Eq  (Succ x) (Succ y));

def nat_ind
( P  : Nat -> Prop,
  f  : (P Zero),
  fn : (Πn:Nat.(P n) -> (P (Succ n))),
   n : Nat 
) :  (P n)  
{
   <(λn:Nat.(P n))> 
   match n  with {
      Zero  => f;
      Succ  => (λn:Nat. (fn n (nat_ind P f fn n)))
   }   
};

def add(x:Nat,y:Nat) : Nat {
  <(λn:Nat.Nat)>
  match x with {
     Zero  => y;
     (Succ z)  => (λz:Nat. (Succ (add z y)))
  }
};

def add_zero(n:Nat) : (Eq (add Zero n) n) {
   (Eq_Rfl n)
};






def add_x_zero(x:Nat): (Eq (add x Zero) x) {
  ((nat_ind
   (λn:Nat.(Eq (add n Zero) n))
   (Eq_Rfl Zero)
   (λn:Nat.(λIH:(Eq (add n Zero) n).
      (Eq_Succ (add n Zero) n IH))))
   x)
};

def plus_x_zero(x:Nat) : (Plus x Zero x) {
  (nat_ind
         (λn:Nat.(Plus n Zero n))
   (Plus_Zero Zero)
   (λy:Nat.(λIH:(Plus y Zero y). (Plus_Succ y Zero y IH)))
   x)
}
\end{lstlisting}

\end{document}